\pdfoutput=1
\RequirePackage{ifpdf}
\ifpdf 
\documentclass[pdftex]{sigma}
\else
\documentclass{sigma}
\fi
\usepackage[all,cmtip]{xy}
\usepackage{slashed}

\newcommand{\starcom}[2]{[#1\stackrel{\star}{,}#2]}
\newcommand{\staranti}[2]{\{#1\stackrel{\star}{,}#2\}}
\newcommand{\ip}[2]{\langle #1, #2\rangle}

\newtheorem{theo}{Theorem}[section]
\newtheorem{lem}[theo]{Lemma}
\newtheorem{propo}[theo]{Proposition}
{\theoremstyle{definition}

\newtheorem{defi}[theo]{Definition}
}

\numberwithin{equation}{section}

\begin{document}

\allowdisplaybreaks

\renewcommand{\thefootnote}{$\star$}

\renewcommand{\PaperNumber}{080}

\FirstPageHeading

\ShortArticleName{Dirac Operators on Noncommutative Curved Spacetimes}

\ArticleName{Dirac Operators on Noncommutative\\
Curved Spacetimes\footnote{This paper is a~contribution to the Special Issue on Deformations of Space-Time
and its Symmetries. The full collection is available at \href{http://www.emis.de/journals/SIGMA/space-time.html}
{http://www.emis.de/journals/SIGMA/space-time.html}}}

\Author{Alexander SCHENKEL~$^\dag$ and Christoph F.~UHLEMANN~$^\ddag$}

\AuthorNameForHeading{A.~Schenkel and C.F.~Uhlemann}

\Address{$^\dag$~Fachgruppe Mathematik, Bergische Universit\"at Wuppertal,\\
\hphantom{$^\dag$}~Gau\ss stra\ss e 20, 42119 Wuppertal, Germany}
\EmailD{\href{mailto:schenkel@math.uni-wuppertal.de}{schenkel@math.uni-wuppertal.de}}

\Address{$^\ddag$~Department of Physics, University of Washington, Seattle, WA 98195-1560, USA}
\EmailD{\href{mailto:uhlemann@uw.edu}{uhlemann@uw.edu}}

\ArticleDates{Received August 09, 2013, in f\/inal form December 11, 2013; Published online December 15, 2013}

\vspace{-1mm}

\Abstract{We study the notion of a~Dirac operator in the framework of twist-deformed noncommutative
geometry.
We provide a~number of well-motivated candidate constructions and propose a~minimal set of axioms that
a~noncommutative Dirac operator should satisfy.
These criteria turn out to be restrictive, but they do not f\/ix a~unique construction: two of our
operators generally satisfy the axioms, and we provide an explicit example where they are inequivalent.
For highly symmetric spacetimes with Drinfeld twists constructed from suf\/f\/iciently many Killing vector
f\/ields, all of our operators coincide.
For general noncommutative curved spacetimes we f\/ind that demanding formal self-adjointness as an
additional condition singles out a~preferred choice among our candidates.
Based on this noncommutative Dirac operator we construct a~quantum f\/ield theory of Dirac f\/ields.
In the last part we study noncommutative Dirac operators on deformed Minkowski and AdS spacetimes as
explicit examples.}


\Classification{81T75; 81T20; 83C65}

\renewcommand{\thefootnote}{\arabic{footnote}}
\setcounter{footnote}{0}

\vspace{-2mm}

\section{Introduction and summary}

Noncommutative geometry has long been of interest from a~purely mathematical perspective as a~natural
generalization of ordinary dif\/ferential geometry.
It is also of crucial interest from a~physical perspective, since it generically plays a~role when the
principles of quantum mechanics are combined with those of general relativity~\cite{Doplicher1,Doplicher2}.
In both contexts, Dirac operators are of major importance: they are relevant for structural questions in
noncommutative geometry~\cite{Connes} and essential for the description of fermionic f\/ields in models for
high-energy physics.
In this article we focus on the latter point and study Dirac operators as equation of motion operators for
Dirac f\/ields on noncommutative spacetimes.
The noncommutative geometries which we are going to consider are obtained by formal deformation
quantization of smooth manifolds via Abelian Drinfeld twists.
Since we are interested in generic curved spacetimes, we will not assume compatibility conditions between
the geometry and the twist.
In particular, we do not restrict ourselves to twists generated solely by Killing vector f\/ields, as these
are just not available on generic spacetimes.

The construction of Dirac operators in our generic setting turns out to be much more involved than in the
highly symmetric setup with twists constructed solely from Killing vector f\/ields.
As shown in~\cite{Connes:2000tj}, the classical Dirac operator is in this special case also a~valid Dirac
operator on the noncommutative manifold.
In the generic case, however, there are several natural deformations of the classical Dirac operator, which
at f\/irst seem equally well motivated and are not obviously equivalent.
It is thus not clear which operator we should choose, and that state of af\/fairs certainly is not
satisfactory.
To improve on it, we will propose an abstract characterization of a~Dirac operator on noncommutative curved
spacetimes, in terms of a~minimal set of axioms.
Namely, it should be a~dif\/ferential operator of f\/irst order in a~sense appropriate for noncommutative
geometry, it should be constructed from geometric objects like the spin connection and the vielbein alone,
and it should have the correct classical limit.
We can then study to which extent the various explicit constructions realize these properties, and under
which circumstances they turn out to be equivalent.

We f\/ind that the minimal set of axioms does in general not uniquely select one of the
constructions that we are going to present.
That is, we show that two of our operators satisfy the axioms and turn out to be inequivalent.
As is to be expected, the classical Dirac operator is not among them.
Restricting then to more special classes of deformations, we f\/ind that for twists constructed from
suf\/f\/iciently many Killing vector f\/ields (semi-Killing twists, which are important for studying
solutions of the noncommutative Einstein equations~\cite{Aschieri:2009qh,Ohl:2009pv, Schupp:2009pt}), our
candidates for Dirac operators all agree and f\/it into our general characterization of noncommutative
Dirac operators.
The freedom in choosing a~Dirac operator is thus reduced drastically in this special class of deformations.
Furthermore, as one is thus free to choose the technically most convenient def\/inition, this can simplify
explicit calculations considerably.
The classical Dirac operator meets our axioms for noncommutative Dirac operators only when restricting to
actual Killing twists, and it then coincides with all the deformed constructions.
Turning back to the general case of twists which do not necessarily involve Killing vector f\/ields, the
question remains which of the (e.g.~inequivalent) noncommutative Dirac operators one should prefer.
Having in mind the construction of noncommutative quantum f\/ield theories, it is natural to also demand
formal self-adjointness with respect to a~suitable inner product.
We will f\/ind that this requirement indeed singles out a~preferred choice among our candidates, and we
will outline the construction of a~quantum f\/ield theory of noncommutative Dirac f\/ields.
A natural next step would be to aim for a~complete classif\/ication of noncommutative Dirac operators
satisfying our axioms, which we leave for future work.

\looseness=-1
The outline of this paper is as follows: For the coupling of Dirac f\/ields to the noncommutative
background geometry we employ techniques of twist-deformed noncommutative geometry and noncommutative
vielbein gravity~\cite{Aschieri:2009ky}, which we review in Section~\ref{sec:preliminaries}.
In Section~\ref{sec:definitionDirac} we discuss three well-motivated deformations of the classical
Dirac operator, and present our minimal set of \mbox{axioms} for Dirac operators in the noncommutative setting.
We then show that two of the proposed operators generally satisfy these axioms.
In Section~\ref{sec:non-uniqueness} we show that for generic noncommutative curved spacetimes the
two noncommutative Dirac operators are inequivalent, and that adding formal self-adjointness as an
additional condition selects a~unique Dirac operator, at least among the examples we have provided.
Furthermore, restricting to a~special class of deformations given by semi-Killing twists we show that the
ambiguities in def\/ining noncommutative Dirac operators disappear in these highly symmetric models, where
the twist contains suf\/f\/iciently many Killing vector f\/ields.
In Section~\ref{sec:quantization} we outline the construction of noncommutative Dirac quantum f\/ield theories.
To illustrate our constructions we provide in Section~\ref{sec:examples} explicit examples and
formulas for noncommutative Dirac operators on spacetimes of physical interest.

\section{Preliminaries}\label{sec:preliminaries}

In the following we review techniques from deformation quantization of smooth manifolds by Abelian Drinfeld
twists and the framework of noncommutative vielbein gravity, as far as they will be relevant for the main part.

\subsection{Twist-deformed noncommutative geometry}

Let $M$ be a~$D$-dimensional manifold.
The noncommutative geometries that we shall consider are those which arise as deformations of $M$ by an
Abelian Drinfeld twist
\begin{gather}
\label{eqn:twist}
\mathcal{F}:=e^{-\frac{i\lambda}{2}\Theta^{\alpha\beta}X_\alpha\otimes X_\beta},
\end{gather}
where $\Theta^{\alpha\beta}$ is an antisymmetric, real and constant matrix (not necessarily of rank $D$)
and $X_\alpha$ are mutually commuting real vector f\/ields on $M$, i.e.\ $[X_\alpha,X_\beta]=0$ for all
$\alpha$, $\beta$.\footnote{The reason for restricting to Abelian Drinfeld twists is the validity of the
graded cyclicity property~\eqref{eqn:gradedcyc}, which does not hold true for generic Drinfeld twists.} The deformation parameter $\lambda$ is assumed to be inf\/initesimally small, i.e.\ we work in formal
deformation quantization.
In this setup a~formal power series extension $\mathbb{C}[[\lambda]]$ of the complex numbers, as well as of
all vector spaces, algebras, etc., has to be performed, but for notational simplicity we will suppress the
square brackets $[[\lambda]]$ denoting these extensions.
We can assume, without loss of generality, that $\Theta^{\alpha\beta}$ is of the canonical (Darboux) form
\begin{gather}\label{eqn:canonical}
\Theta=
\begin{pmatrix}
0&1&0&0&\cdots
\\
-1&0&0&0&\cdots
\\
0&0&0&1&\cdots
\\
0&0&-1&0&\cdots
\\
\vdots&\vdots&\vdots&\vdots&\ddots
\end{pmatrix}.
\end{gather}

It is convenient to introduce the following notation.

\begin{defi}
A {\it twisted manifold} is a~pair $(M,\mathcal{F})$, where $M$ is a~$D$-dimensional manifold and~$\mathcal{F}$ is an Abelian Drinfeld twist, cf.~\eqref{eqn:twist}.
\end{defi}

On any twisted manifold $(M,\mathcal{F})$ we can develop a~canonical noncommutative dif\/ferential geometry.
As a~f\/irst step, consider $C^\infty(M)$, the space of all complex-valued smooth functions on $M$.
By using the twist $\mathcal{F}$, we can equip this space with an associative and noncommutative product
(the $\star$-product)
\begin{gather*}
f\star g:=\mu\big(\mathcal{F}^{-1}f\otimes g\big)=fg+\frac{i\lambda}{2}\Theta^{\alpha\beta}
X_\alpha(f)X_\beta(g)+\cdots,
\end{gather*}
where $\mu$ is the usual point-wise product and the action of the vector f\/ields $X_\alpha$ on the
func\-tions~$f$,~$g$ is via the (Lie) derivative.

Moreover, the de Rham calculus $(\Omega^\bullet(M):= \bigoplus_{n=0}^D
\Omega^n(M),\wedge,\mathrm{d})$ on $M$ can be deformed by~$\mathcal{F}$ into a~dif\/ferential calculus
on the $\star$-product algebra $(C^\infty(M),\star)$.
Explicitly, we def\/ine the $\wedge_\star$-product
\begin{gather*}
\omega\wedge_\star\tau:=\wedge\big(\mathcal{F}^{-1}\omega\otimes\tau\big)
=\omega\wedge\tau+\frac{i\lambda}{2}\Theta^{\alpha\beta}\mathcal{L}_{X_\alpha}(\omega)\wedge\mathcal{L}_{X_\beta}(\tau)
+\cdots,
\end{gather*}
where the action of the vector f\/ields $X_\alpha$ on the dif\/ferential forms $\omega$, $\tau$ is via the Lie
derivative.
The undeformed dif\/ferential $\mathrm{d}$ satisf\/ies the graded Leibniz rule with respect to the
$\wedge_\star$-product, i.e.\ $\mathrm{d}(\omega\wedge_\star \tau) = (\mathrm{d}\omega)\wedge_\star \tau +
(-1)^{\vert\omega\vert}\omega\wedge_\star(\mathrm{d}\tau)$ with $\vert \omega\vert$ denoting the degree
of $\omega$, and hence $(\Omega^\bullet(M),\wedge_\star,\mathrm{d})$ is a~dif\/ferential calculus
over $(C^\infty(M),\star)$.
We extend the involution $\ast$ on $(C^\infty(M),\star)$, which is given by point-wise complex
conjugation, to a~graded involution on $(\Omega^\bullet(M),\wedge_\star,\mathrm{d})$ by applying
the rules $(\omega\wedge_\star\tau)^\ast =(-1)^{\vert\omega\vert\,\vert\tau\vert} \tau^\ast \wedge_\star
\omega^\ast$ and $(\mathrm{d} \omega)^\ast = \mathrm{d}(\omega^\ast)$.
The undeformed integral $\int_M: \Omega^D(M) \to \mathbb{C}$ satisf\/ies the graded cyclicity property.
This means that, for all $\omega,\tau \in \Omega^\bullet(M)$ with compact overlapping support and such that
$\vert\omega\vert + \vert\tau\vert =D$,
\begin{gather}
\label{eqn:gradedcyc}
\int_M\omega\wedge_\star\tau=\int_M\omega\wedge\tau=(-1)^{\vert\omega\vert\,\vert\tau\vert}
\int_M\tau\wedge_\star\omega.
\end{gather}

Finally, we deform the contraction operator $\iota$ (interior product) between vector f\/ields and
one-forms on $M$ by the twist $\mathcal{F}$.
The resulting $\star$-contraction operator $\iota^\star$ is given by (cf.~\cite{Aschieri:2005zs})
\begin{gather}
\label{eqn:contractioniota}
\iota^\star_v(\omega):=\iota\big(\mathcal{F}^{-1}v\otimes\omega\big)=\iota_v(\omega)+\frac{i\lambda}{2}
\Theta^{\alpha\beta}\iota_{\mathcal{L}_{X_\alpha}(v)}\big(\mathcal{L}_{X_\beta}(\omega)\big)+\cdots,
\end{gather}
where again the vector f\/ields $X_\alpha$ act via the Lie derivative on vector f\/ields~$v$ and one-forms~$\omega$.

This completes our snapshot review of twist-deformed noncommutative geometry and we refer the reader
to~\cite{Schenkel:2012zu} for a~more detailed discussion.

\subsection{Noncommutative vielbein gravity}

The explicit constructions and examples of the main part will involve spacetimes of dimension~$2$ and~$4$,
and we therefore give the formalism of noncommutative vielbein gravity for both cases in the following.
Throughout, $(M,\mathcal{F})$ will denote a~twisted manifold of appropriate dimension.
Following~\cite{Aschieri:2009ky}, we describe the noncommutative gravitational f\/ield by a~noncommutative
vielbein f\/ield $V$ and a~noncommutative spin connection $\Omega$.
Both are Clif\/ford-algebra valued one-forms.
Our gamma-matrix conventions are collected for easy reference in Appendix~\ref{app:gamma-conv}.
We say that the noncommutative spin connection is $\star$-torsion free if $0=\mathrm{d}_\Omega V:=
\mathrm{d} V - \staranti{\Omega}{V} $, where $\staranti{\Omega}{V} = \Omega\wedge_\star V + V\wedge_\star
\Omega$ is the $\star$-anticommutator.
The $\star$-torsion constraint is part of the equations of motion of noncommutative vielbein
gravity~\cite{Aschieri:2009ky}.
For reasons of generality, we do not assume the $\star$-torsion constraint for our general constructions
and we shall clearly indicate at which later step it is used.

{\bf $\boldsymbol{D=4}$.} We can expand $V$ and $\Omega$ in terms of the gamma-matrix basis
$\{1,\gamma_5,\gamma_a,\gamma_a\gamma_5,\gamma_{ab}\}$ of the $4$-dimensional Clif\/ford algebra as
\begin{gather}
\label{eqn:component}
V=V^a\gamma_a+\tilde V^a\gamma_a\gamma_5
,
\qquad
\Omega=\frac{1}{4}\omega^{ab}\gamma_{ab}+i\omega1+\tilde\omega\gamma_5.
\end{gather}
We further demand the reality conditions $V^\dagger = \gamma_0 V\gamma_0$ and $\Omega^\dagger =
-\gamma_0\Omega\gamma_0$.
Notice that noncommutative $D=4$ vielbein gravity contains more f\/ields than its commutative counterpart,
where $\tilde V^a =\omega = \tilde \omega = 0$.
The reason is that ${\rm SL}(2,\mathbb{C})$ (Lorentz) \mbox{$\star$-gauge} transformations do not close and have to be
extended to ${\rm GL}(2,\mathbb{C})$ \mbox{$\star$-gauge} transformations.
The \mbox{$\star$-gauge} transformations act on $V$ and $\Omega$ by
\begin{gather}
\label{eqn:gaugetrafos}
\delta_\epsilon V=\starcom{\epsilon}{V},
\qquad
\delta_\epsilon\Omega=\mathrm{d}\epsilon+\starcom{\epsilon}{\Omega},
\end{gather}
where $\epsilon = \frac{1}{4}\epsilon^{ab}\gamma_{ab} + i\varepsilon 1 + \tilde\varepsilon
\gamma_5$ is a~Clif\/ford algebra valued function and $\starcom{\epsilon}{V}:= \epsilon \star V - V\star
\epsilon$ is the $\star$-commutator.
We impose the reality condition $\epsilon^\dagger = -\gamma_0\epsilon\gamma_0$.

As in~\cite{Aschieri:2009ky} we require that $\tilde V^a \vert_{\lambda=0}=\omega\vert_{\lambda=0} = \tilde
\omega\vert_{\lambda=0} = 0$, such that the commutative limit yields a~usual commutative ${\rm SL}(2,\mathbb{C})$
vierbein and spin connection.
We shall use the notation $(V_{(0)},\Omega_{(0)}):= (V\vert_{\lambda=0}^{},\Omega\vert_{\lambda=0}^{})$.
\begin{defi}
Let $(M,\mathcal{F})$ be a~$4$-dimensional twisted manifold.
A {\it noncommutative Cartan geometry} on $(M,\mathcal{F})$ is a~pair of Clif\/ford algebra valued
one-forms $(V,\Omega)$, satisfying the expansion~\eqref{eqn:component}, the reality conditions $V^\dagger =
\gamma_0 V\gamma_0$, $\Omega^\dagger = -\gamma_0\Omega\gamma_0$ and the limit $\tilde V^a
\vert_{\lambda=0}=\omega\vert_{\lambda=0} = \tilde \omega\vert_{\lambda=0} = 0$.
\end{defi}

Let us now consider Dirac f\/ields, i.e.\ functions $\psi\in C^\infty(M,\mathbb{C}^4)$ valued in the
fundamental representation of the Clif\/ford algebra.
We denote the Dirac adjoint by $\overline{\psi}:= \psi^\dagger\gamma_0$.
The \mbox{$\star$-gauge} transformations act on $\psi$ and $\overline{\psi}$ by $\delta_\epsilon \psi =
\epsilon\star \psi$ and $\delta_\epsilon \overline{\psi} = - \overline{\psi}\star \epsilon $, respectively.
Notice that the matrix $\psi \star \overline{\psi}$ transforms in the adjoint representation,
$\delta_\epsilon (\psi\star\overline{\psi}) = \starcom{\epsilon}{ \psi\star\overline{\psi}}$.
For all Dirac f\/ields~$\psi_1$,~$\psi_2$ with compact overlapping support we def\/ine the inner product{\samepage
\begin{gather}
\label{eqn:innerproduct}
\ip{\psi_1}{\psi_2}:=i\int_M\mathrm{Tr}\big(\psi_2\star\overline{\psi_1}
\star V\wedge_\star V\wedge_\star V\wedge_\star V\gamma_5\big),
\end{gather}
which is \mbox{$\star$-gauge} invariant due to~\eqref{eqn:gradedcyc},~\eqref{eqn:gaugetrafos} and the cyclicity of
the matrix trace $\mathrm{Tr}$.}

\begin{lem}
Let $(V,\Omega)$ be a~noncommutative Cartan geometry on a~$4$-dimensional twisted mani\-fold
$(M,\mathcal{F})$.
Then the inner product~\eqref{eqn:innerproduct} is hermitian, it reduces to the canonical commutative one
for $\lambda=0$ and it is non-degenerate, i.e.:
\begin{itemize}\itemsep=0pt
\item[$a)$] $\ip{\psi_1}{\psi_2}^\ast = \ip{\psi_2}{\psi_1} $.
\item[$b)$] $\ip{\psi_1}{\psi_2} = \int_M
\overline{\psi_1} \psi_2\mathrm{vol}_{(0)} + \mathcal{O}(\lambda)$, where $\mathrm{vol}_{(0)} =
V_{(0)}^a\wedge V_{(0)}^b\wedge V_{(0)}^c\wedge V_{(0)}^d\epsilon_{abcd}$.
\item[$c)$] If
$\ip{\psi_1}{\psi_2} =0$ for all $\psi_2$, then $\psi_1 =0$.
\end{itemize}
\end{lem}

\begin{proof}
We show $a)$ by the following short calculation
\begin{gather*}
\ip{\psi_1}{\psi_2}^\ast=-i\int_M\mathrm{Tr}
\big(\gamma_5^\dagger V^\dagger\wedge_\star V^\dagger\wedge_\star V^\dagger\wedge_\star V^\dagger\star(\psi_2\star\overline{\psi_1})^\dagger\big)
\\
\phantom{\ip{\psi_1}{\psi_2}^\ast}{}
=-i\int_M\mathrm{Tr}\big(\psi_1\star\overline{\psi_2}
\gamma_0\gamma_5\gamma_0\star V\wedge_\star V\wedge_\star V\wedge_\star V\big)
\\
\phantom{\ip{\psi_1}{\psi_2}^\ast}{}
=i\int_M\mathrm{Tr}\big(\psi_1\star\overline{\psi_2}
\star V\wedge_\star V\wedge_\star V\wedge_\star V\gamma_5\big)=\ip{\psi_2}{\psi_1}.
\end{gather*}
In the second equality we have used (graded) cyclicity, the reality condition $V^\dagger = \gamma_0
V\gamma_0$, $(\psi_2\star \overline{\psi_1})^\dagger = \gamma_0\psi_1\star\overline{\psi_2}\gamma_0$,
$\gamma_0^2 =1$ and $\gamma_5^\dagger=\gamma_5$.
In the third equality we have used $\gamma_5 \gamma_0 = -\gamma_0\gamma_5$, $\gamma_0^2=1$ and $\gamma_5 V
= -V\gamma_5$.

To show $b)$ let us set in~\eqref{eqn:innerproduct} $\lambda =0$ and use that $V\vert_{\lambda=0} =
V_{(0)}^a\gamma_a$ (i.e.\ that $\tilde V^a$ vanishes at order $\lambda^0$).
Using further that the antisymmetrized product of 4 gamma-matrices is $\gamma_{[a}
\gamma_b\gamma_c\gamma_{d]} = -i\gamma_5\epsilon_{abcd}$ and that $\gamma_5^2=1$ we obtain the desired
result.

$c)$ is a~consequence of $b)$ and the fact that the classical inner product $\int_M\overline{\psi_1}\psi_2
\mathrm{vol}_{(0)}$ is non-degenerate.
\end{proof}

{\bf $\boldsymbol{D=2}$.} The noncommutative twobein and spin connection have the following expansion in
terms of the gamma-matrix basis $\{1,\gamma_3,\gamma_a\}$ of the $2$-dimensional Clif\/ford algebra
\begin{gather}
\label{eqn:components2D}
V=V^a\gamma_a
,
\qquad
\Omega=\omega\gamma_3+\tilde{\omega}1.
\end{gather}
We def\/ine for $\epsilon = \varepsilon\gamma_3 + \tilde\varepsilon1$ the \mbox{$\star$-gauge} transformations
$\delta_\epsilon V:= \starcom{\epsilon}{V} $ and $\delta_\epsilon\Omega:= \mathrm{d} \epsilon +
\starcom{\epsilon}{\Omega}$.
As in the case of $D=4$, we had to introduce the extra f\/ields $\tilde{\omega}$ and $\tilde\varepsilon$
such that the \mbox{$\star$-gauge} transformations close.
Note, however, that we do not need additional terms in the twobein f\/ield and thus the interpretation of
$V$ as a~soldering form remains valid in $D=2$.
This will facilitate the study of noncommutative Dirac operators in $D=2$, as compared to $D=4$.
We again impose the reality conditions $\epsilon^\dagger = -\gamma_0\epsilon\gamma_0$, $V^\dagger =
\gamma_0V\gamma_0$ and $\Omega^\dagger = -\gamma_0\Omega\gamma_0$.
\begin{defi}
Let $(M,\mathcal{F})$ be a~$2$-dimensional twisted manifold.
A {\it noncommutative Cartan geometry} on $(M,\mathcal{F})$ is a~pair of Clif\/ford algebra valued
one-forms $(V,\Omega)$, satisfying the expansion~\eqref{eqn:components2D}, the reality conditions
$V^\dagger = \gamma_0 V\gamma_0$, $\Omega^\dagger = -\gamma_0\Omega\gamma_0$ and the limit $\tilde
\omega\vert_{\lambda=0} = 0$.
\end{defi}

Let us now consider Dirac f\/ields, which in the case of $D=2$ are functions $\psi\in
C^\infty(M,\mathbb{C}^2)$ with values in the fundamental representation of the $D=2$ Clif\/ford algebra.
The Dirac adjoint is $\overline{\psi}:= \psi^\dagger \gamma_0$ and \mbox{$\star$-gauge} transformations act on
$\psi$ and $\overline{\psi}$ via $\delta_\epsilon \psi:= \epsilon\star \psi$ and $\delta_\epsilon
\overline{\psi} = -\overline{\psi}\star\epsilon$.
We def\/ine in analogy to~\eqref{eqn:innerproduct} a~\mbox{$\star$-gauge} invariant and hermitian inner product
\begin{gather}
\label{eqn:2dinnerproduct}
\ip{\psi_1}{\psi_2}:=\int_M\mathrm{Tr}\big(\psi_2\star\overline{\psi_1}\star V\wedge_\star V\gamma_3\big).
\end{gather}
For $\lambda=0$ we obtain the usual inner product $\ip{\psi_1}{\psi_2}= \int_M
\overline{\psi_1}\psi_2\mathrm{vol}_{(0)} + \mathcal{O}(\lambda)$, since $V\wedge_\star V \gamma_3
\vert_{\lambda=0}= V_{(0)}\wedge V_{(0)}\gamma_3 = V_{(0)}^a\wedge V_{(0)}^b \epsilon_{ab}\gamma_3^2 =
\mathrm{vol}_{(0)}$.

\section{Noncommutative Dirac operators}\label{sec:definitionDirac}

As a~f\/irst step, we will give three explicit candidate def\/initions for noncommutative Dirac operators
in $D=2$ and $D=4$.
These are obtained by using techniques of noncommutative dif\/ferential geometry and twist deformation
quantization.
Depending on taste and point of view, either of them may be seen as a~valid extension of the classical
Dirac operator to the noncommutative setting.
This shows that a~more abstract characterization of noncommutative Dirac operators is needed, and we
develop in the second step what we believe is a~minimal set of axioms for such operators.
This will already rule out the classical Dirac operator along with one of our candidates, and we show that
the remaining two indeed meet our criteria for noncommutative Dirac operators.

\subsection{Explicit candidates}

The following set of candidates for noncommutative Dirac operators should show, how focusing on dif\/ferent
aspects of noncommutative dif\/ferential geometry and twist deformation quantization leads to dif\/ferent
constructions.
It is not meant to be exhaustive.

{\bf The Aschieri--Castellani Dirac operator.} The f\/irst operator we consider is motivated by the
noncommutative $D=4$ Dirac f\/ield action proposed in~\cite{Aschieri:2009ky}, which reads
\begin{gather}
\label{eqn:Diracaction}
S^\mathrm{AC}=-4\int_M\mathrm{Tr}\big((\mathrm{d}_\Omega\psi)\star\overline{\psi}
\wedge_\star V\wedge_\star V\wedge_\star V\gamma_5\big),
\end{gather}
where $\mathrm{d}_\Omega \psi:= \mathrm{d}\psi - \Omega \star \psi$ is the $\star$-covariant dif\/ferential
acting on Dirac f\/ields.
Since the inner product~\eqref{eqn:innerproduct} is non-degenerate, we can def\/ine a~dif\/ferential
operator $\slashed{\mathrm{D}}^{\mathrm{AC}}:C^\infty(M,\mathbb{C}^4)\to C^\infty(M,\mathbb{C}^4)$ by
requiring that, for all $\psi_1$ of compact support,
\begin{gather}
\label{eqn:ACDirac}
\ip{\psi_1}{\slashed{\mathrm{D}}^\mathrm{AC}\psi_2}=-4\int_M\mathrm{Tr}\big((\mathrm{d}
_\Omega\psi_2)\star\overline{\psi_1}\wedge_\star V\wedge_\star V\wedge_\star V\gamma_5\big).
\end{gather}
This yields exactly the equation of motion operator which is obtained by varying the
action~\eqref{eqn:Diracaction} with respect to $\overline{\psi}$.
The construction in $D=2$ is fully analogous.
The action then reads
\begin{gather*}
S^\mathrm{AC}=2i\int_M\mathrm{Tr}\big((\mathrm{d}_\Omega\psi)\star\overline{\psi}\wedge_\star V\gamma_3\big),
\end{gather*}
and since the inner product~\eqref{eqn:2dinnerproduct} is also non-degenerate, we can def\/ine
a~dif\/ferential operator $\slashed{\mathrm{D}}^\mathrm{AC}: C^\infty(M,\mathbb{C}^2)\to
C^\infty(M,\mathbb{C}^2)$ by requiring that, for all $\psi_1$ of compact support,
\begin{gather}
\label{eqn:2dACDirac}
\ip{\psi_1}{\slashed{\mathrm{D}}^\mathrm{AC}\psi_2}=2i\int_M\mathrm{Tr}\big((\mathrm{d}
_\Omega\psi_2)\star\overline{\psi_1}\wedge_\star V\gamma_3\big).
\end{gather}

{\bf The contraction Dirac operator.} For our next operator we shall follow closely the usual
construction of a~Dirac operator on commutative spacetimes, which goes as follows: Let $V_{(0)} =
V_{(0)}^a\gamma_a$ be a~classical vielbein, $\Omega_{(0)}$ a~classical spin connection and let us denote by
$V^{-1}_{(0)}= E_{(0)a}\gamma^a$ the inverse vielbein.
The classical Dirac operator is $\slashed{\mathrm{D}}_{(0)}\psi = i\gamma^a\nabla_{(0)a}\psi =
i\gamma^a (E_{(0)a}(\psi) - \Omega_{(0)a}\psi) $, where we have expressed $\Omega_{(0)}$ in
the vielbein basis $\Omega_{(0)} = V^a\Omega_{(0)a}$.
Notice that this operator can be written in an index-free form $\slashed{\mathrm{D}}_{(0)}\psi =
i\iota_{V^{-1}_{(0)}} (\mathrm{d} \psi - \Omega_{(0)}\psi)$, where $\iota$ is the classical
contraction operator (interior product).

Using the deformed contraction operator $\iota^\star$ between vector f\/ields and one-forms as def\/ined
in~\eqref{eqn:contractioniota}, we generalize the above construction to the noncommutative setting.
For this we def\/ine the $\star$-inverse vielbein $E_a$ by the $\star$-contraction condition
$\iota^\star_{E_{a}}(V^b) = \delta_{a}^b$.
We collect all $E_a$ in the Clif\/ford algebra valued vector f\/ield $V^{-1_\star}:= E_a\gamma^a$.
Following the same strategy as in the classical case, we def\/ine a~dif\/ferential operator $
\slashed{\mathrm{D}}^\mathrm{contr}$ by
\begin{gather}
\label{eqn:2dcontrDirac}
\slashed{\mathrm{D}}^\mathrm{contr}\psi:=i\iota^\star_{V^{-1_\star}}\big(\mathrm{d}
_\Omega\psi\big)=i\gamma^a\iota^{\star}_{E_a}\big(\mathrm{d}_\Omega\psi\big).
\end{gather}
The construction outlined above is valid as it stands in $D=2$.
For $D=4$ the following remark is in order: As seen in~\eqref{eqn:component}, the noncommutative vierbein
f\/ield has an extra f\/ield $\widetilde{V}^a$.
This implies that $V$ is locally a~$4\times 8$-matrix and hence there is no unique $\star$-inverse
$V^{-1_\star}$.
Since invertibility of the vielbein is essential in classical vielbein gravity, this may be seen as
a~shortcoming of the noncommutative $D=4$ vielbein gravity formulated in~\cite{Aschieri:2009ky}.
There have been attempts to overcome this issue by using Seiberg--Witten maps~\cite{Aschieri:2011ng},
which, however, obscure the noncommutative dif\/ferential geometry and in practice require an expansion in
the deformation parameter to some f\/ixed order, so they are not convenient for our purpose.
We shall instead take the following approach: We restrict the class of allowed noncommutative Cartan
geometries $(V,\Omega)$ to those satisfying $V=V^a\gamma_a$.
We are aware that this restriction is not invariant under \mbox{$\star$-gauge} transformations (in $D=4$) and
hence it is not convenient for {\it dynamical} noncommutative $D=4$ vielbein gravity.
However, for a~given {\it fixed} noncommutative Cartan geometry $(V,\Omega)$ it certainly makes sense to
assume a~$V$ of this special form.

{\bf The deformed Dirac operator.} The last noncommutative Dirac operator is motivated by the
framework of Connes for noncommutative spin geometry~\cite{Connes}, where the Dirac operator enters as
a~fundamental degree of freedom of the theory.
It is obtained by deforming the classical Dirac operator $\slashed{\mathrm{D}}_{(0)}$ via the techniques
developed in~\cite{Aschieri:2012ii}.
More precisely, denoting the inverse twist by $\mathcal{F}^{-1}=\bar f^\alpha\otimes \bar f_\alpha$, we
def\/ine the deformed Dirac operator by applying the deformation map constructed in~\cite{Aschieri:2012ii}
\begin{gather*}
\slashed{\mathrm{D}}^{\mathcal{F}}\psi:=\big(\bar f^\alpha\blacktriangleright\slashed{\mathrm{D}}_{(0)}
\big)\bar f_\alpha(\psi)=\slashed{\mathrm{D}}_{(0)}\psi+\frac{i\lambda}{2}\Theta^{\alpha\beta}
\big(X_\alpha\blacktriangleright\slashed{\mathrm{D}}_{(0)}\big)X_\beta(\psi)+\cdots,
\end{gather*}
where $X_\alpha \blacktriangleright \slashed{\mathrm{D}}_{(0)}:= X_\alpha \circ \slashed{\mathrm{D}}_{(0)}
- \slashed{\mathrm{D}}_{(0)}\circ X_\alpha$ is the adjoint action.

\subsection{Abstract consideration}

The variety of dif\/ferent generalizations of the classical Dirac operator to the noncommutative setting
provided above clearly calls for a~more precise def\/inition of what we actually mean by a~noncommutative
Dirac operator.
As a~reasonable starting point, we are looking for linear dif\/ferential operators $\slashed{\mathrm{D}}:
C^\infty(M,\mathbb{C}^{N})\to C^\infty(M,\mathbb{C}^N)$ acting on Dirac f\/ields $\psi\in
C^\infty(M,\mathbb{C}^N)$ ($N$ is the dimension of the fundamental representation of the Clif\/ford
algebra), subject to certain conditions generalizing the properties of the classical Dirac operator.
Thus, we naturally start by generalizing some relevant notions to the noncommutative setting, beginning
with the notion of a~f\/irst-order dif\/ferential operator.
\begin{defi}
Let $(M,\mathcal{F})$ be a~twisted manifold.
A dif\/ferential operator $\slashed{\mathrm{D}}: C^\infty(M,\mathbb{C}^N)\to C^\infty(M,\mathbb{C}^N)$ is
called a~{\it first-order noncommutative differential operator}, if for all $\psi\!\in\!
C^\infty(M,\mathbb{C}^N)$ and $a\in C^\infty(M)$,
\begin{gather}
\label{eqn:firstorder}
\slashed{\mathrm{D}}(\psi\star a)=\slashed{\mathrm{D}}(\psi)\star a+\iota_{Q_\psi}^\star(\mathrm{d}a),
\end{gather}
where $Q_\psi$ is a~spinor-valued vector f\/ield on $M$ and the $\star$-contraction is def\/ined
in~\eqref{eqn:contractioniota}.
\end{defi}

For $\lambda=0$ we obtain from~\eqref{eqn:firstorder} the usual Leibniz rule property of a~f\/irst-order
dif\/ferential ope\-rator.
Written in local coordinates it reads $\slashed{\mathrm{D}}(\psi a) = \slashed{\mathrm{D}}(\psi)a +
Q_\psi^\mu\partial_\mu a$.
Hence,~\eqref{eqn:firstorder} promotes this property to the realm of twisted manifolds.
Note that a~f\/irst-order noncommutative dif\/ferential operator is not necessarily a~f\/irst-order
dif\/ferential operator in the usual sense (cf.~the examples in the sections below).
It can and in general must contain higher order derivatives, but these are restricted by the form of the
twist $\mathcal{F}$.

The second notion we want to generalize aims to capture more of the essence of the classical Dirac operator.
Namely, that it is constructed from purely geometric data in a~natural way.
We will formalize the requirement that a~noncommutative Dirac operator should be constructed only from the
data of the noncommutative Cartan geometry $(V,\Omega)$ and the twisted manifold $(M,\mathcal{F})$ in
a~geometric (natural) way as follows\footnote{This can be made more precise in a~category theoretical
framework for noncommutative Cartan geometries on twisted manifolds, where a~natural dif\/ferential
operator could be def\/ined in terms of a~natural transformation between the section functors of the Dirac
bundles.}:
\begin{defi}
Let $(V,\Omega)$ be a~noncommutative Cartan geometry on a~twisted manifold $(M,\mathcal{F})$.
A dif\/ferential operator $\slashed{\mathrm{D}}: C^\infty(M,\mathbb{C}^N)\to C^\infty(M,\mathbb{C}^N)$ is
called a~{\it geometric noncommutative differential operator} if it is constructed from the
noncommutative vielbein $V$ and the $\star$-covariant dif\/ferential $\mathrm{d}_\Omega$ in terms of the
operations of twisted noncommutative geometry.
\end{defi}

We can now state our general def\/inition for noncommutative Dirac operators, which combines the properties
introduced above with the natural demand that the standard Dirac operator should be recovered in the
commutative limit:
\begin{defi}
\label{defi:Dirac}
A {\it noncommutative Dirac operator} on a~noncommutative Cartan geometry $(V,\Omega)$ over a~twisted
manifold $(M,\mathcal{F})$ is a~dif\/ferential operator $\slashed{\mathrm{D}}: C^\infty(M,\mathbb{C}^N)\to
C^\infty(M,\mathbb{C}^N)$, such that
\begin{itemize}\itemsep=0pt
\item[1)]
$\slashed{\mathrm{D}}$ is a~f\/irst-order noncommutative dif\/ferential operator,
\item[2)]
$\slashed{\mathrm{D}}$ is a~geometric noncommutative dif\/ferential operator,
\item[3)]
$\slashed{\mathrm{D}}$ reproduces the classical Dirac operator $\slashed{\mathrm{D}}_{(0)}$ corresponding
to $(V_{(0)}, \Omega_{(0)})$ for $\lambda=0$.
\end{itemize}
\end{defi}
Having stated this def\/inition, the f\/irst two questions one could ask are, f\/irstly, whether there are
noncommutative Dirac operators in this sense at all, and, secondly, whether the requirements are possibly
trivial altogether.
In the remaining part of this subsection we will answer these two questions, focusing on $D=2$ and $D=4$.

The f\/irst question is easily answered by providing an explicit construction which satisf\/ies the demands
of Def\/inition~\ref{defi:Dirac}:
\begin{propo}
Let $(M,\mathcal{F})$ be any $2$- or $4$-dimensional twisted manifold and $(V,\Omega)$ any noncommutative
Cartan geometry.
Then the operator $\slashed{\mathrm{D}}^\mathrm{AC}$ defined in~\eqref{eqn:2dACDirac}
and~\eqref{eqn:ACDirac} is a~noncommutative Dirac operator according to Definition~{\rm \ref{defi:Dirac}}.
\end{propo}

\begin{proof}
We give the proof for $D=2$, and note that it follows analogously for $D=4$.
We have to check the three conditions in Def\/inition~\ref{defi:Dirac}.
For property~1) let us evaluate the following inner product
\begin{gather*}
\ip{\psi_1}{\slashed{\mathrm{D}}^\mathrm{AC}(\psi_2\star a)}=2i\int_M\mathrm{Tr}\big((\mathrm{d}
_\Omega(\psi_2\star a))\star\overline{\psi_1}\wedge_\star V\gamma_3\big)
\\
\phantom{\ip{\psi_1}{\slashed{\mathrm{D}}^\mathrm{AC}(\psi_2\star a)}}{}
=2i\int_M\mathrm{Tr}\big(((\mathrm{d}_\Omega\psi_2)\star a+\psi_2\star\mathrm{d}
a)\star\overline{\psi_1}\wedge_\star V\gamma_3\big)
\\
\phantom{\ip{\psi_1}{\slashed{\mathrm{D}}^\mathrm{AC}(\psi_2\star a)}}{}
=\ip{\psi_1}{\slashed{\mathrm{D}}^\mathrm{AC}(\psi_2)\star a}+2i\int_M\mathrm{Tr}
\big(\psi_2\star\mathrm{d}a\star\overline{\psi_1}\wedge_\star V\gamma_3\big).
\end{gather*}
The proof would follow if we could show that the condition
\begin{gather}
\label{eqn:tmp2dACDiracinverse}
V\wedge_\star V\gamma_3\star\iota^\star_{Q_{\psi_2}}(\omega)=-2iV\gamma_3\star\psi_2\wedge_\star\omega,
\qquad
\text{for all}
\quad
\omega\in\Omega^1(M),
\end{gather}
def\/ines a~unique spinor-valued vector f\/ield $Q_{\psi_2}$ on $M$.
This is indeed the case by the following argument: Notice that $V\wedge_\star V\gamma_3$ is equal to the
classical volume form $\mathrm{vol}_{(0)}$ at order $\lambda^0$.
Since this form is non-degenerate, the condition~\eqref{eqn:tmp2dACDiracinverse} determines a~unique
spinor-valued function $\iota_{Q_{\psi_2}}^\star(\omega)$, for any $\omega\in \Omega^1(M)$.
The operation $\iota^\star_{Q_{\psi_2}}$ is right linear under the $\star$-multiplication by $C^\infty(M)$,
for all $\omega\in \Omega^1(M)$ and $a\in C^\infty(M)$,
\begin{gather*}
V\wedge_\star V\gamma_3\star\iota^\star_{Q_{\psi_2}}
(\omega\star a)=-2iV\gamma_3\star\psi_2\wedge_\star\omega\star a
=V\wedge_\star V\gamma_3\star\iota^\star_{Q_{\psi_2}}(\omega)\star a.
\end{gather*}
Since the space of vector f\/ields is exactly the dual module of the module of one-forms, $Q_{\psi_2}$ is
a~spinor-valued vector f\/ield.

Property 2) is clear: We have used only $\star$-products, $\wedge_\star$-products, integrals $\int_M$,
vielbeins $V$ and $\star$-covariant dif\/ferentials $\mathrm{d}_\Omega$ in order to def\/ine
$\slashed{\mathrm{D}}^\mathrm{AC}$.

To prove property 3) we consider~\eqref{eqn:2dACDirac} at $\lambda=0$.
We expand $\mathrm{d}_\Omega\psi_2\vert_{\lambda=0}^{}$ in the twobein basis $V_{(0)}^a$,
i.e.\ $\mathrm{d}_\Omega\psi_2\vert_{\lambda=0}^{} = V_{(0)}^a (E_{(0)a}(\psi_2) - \Omega_{(0)a}
\psi_2)$, where $E_{(0)a}$ is the inverse of $V_{(0)}^a$, which is a~vector f\/ield.
We obtain
\begin{gather*}
\ip{\psi_1}{\slashed{\mathrm{D}}^\mathrm{AC}\psi_2}\vert_{\lambda=0}
=2i\int_M\mathrm{Tr}
\left(V_{(0)}^b\big(E_{(0)b}(\psi_2)-\Omega_{(0)b}\psi_2\big)\overline{\psi_1}\wedge V_{(0)}
^a\gamma_a\gamma_3\right)
\\
\phantom{\ip{\psi_1}{\slashed{\mathrm{D}}^\mathrm{AC}\psi_2}\vert_{\lambda=0}}{}
=2i\int_M\overline{\psi_1}\epsilon_{ac}\gamma^c\big(E_{(0)b}(\psi_2)-\Omega_{(0)b}
\psi_2\big)V_{(0)}^a\wedge V_{(0)}^b
\\
\phantom{\ip{\psi_1}{\slashed{\mathrm{D}}^\mathrm{AC}\psi_2}\vert_{\lambda=0}}{}
=\int_{M}\overline{\psi_1}i\gamma^a\big(E_{(0)a}(\psi_2)-\Omega_{(0)a}\psi_2\big)\mathrm{vol}
_{(0)}=\int_M\overline{\psi_1}\big(\slashed{\mathrm{D}}_{(0)}\psi_2\big)\mathrm{vol}_{(0)}.\tag*{\qed}
\end{gather*}
\renewcommand{\qed}{}
\end{proof}
Bearing in mind the remark on the $D=4$ case below Equation~\eqref{eqn:2dcontrDirac}, also the contraction
Dirac operator is valid in the sense of Def\/inition~\ref{defi:Dirac}:
\begin{propo}
Let $(M,\mathcal{F})$ be a~twisted manifold of dimension $2$ or $4$, and $(V,\Omega)$ a~noncommutative
Cartan geometry, such that $V= V^a\gamma_a$.
Then the operator $\slashed{\mathrm{D}}^\mathrm{contr}$ defined in~\eqref{eqn:2dcontrDirac} is
a~noncommutative Dirac operator according to Definition~{\rm \ref{defi:Dirac}}.
\end{propo}
\begin{proof}
We have to check the three conditions in Def\/inition~\ref{defi:Dirac}.
Property~1) follows from a~short calculation
\begin{gather*}
\slashed{\mathrm{D}}^\mathrm{contr}(\psi\star a)
=i\iota^\star_{V^{-1_\star}}\big((\mathrm{d}_\Omega\psi)\star a+\psi\star\mathrm{d}a\big)
=\slashed{\mathrm{D}}^\mathrm{contr}(\psi)\star a+\iota_{Q_{\psi}}^\star(\mathrm{d}a),
\end{gather*}
where $Q_\psi = iV^{-1_\star} \star \psi$.

Property 2) is clear: We have only used $V$, $\mathrm{d}_\Omega$, the $\star$-inverse $V^{-1_\star}$
(def\/ined via $\iota_\star$) and $\iota_\star$ to construct $\slashed{\mathrm{D}}^\mathrm{contr}$.
Furthermore, property 3) is a~consequence of the fact that all operations
entering~\eqref{eqn:2dcontrDirac} reduce for $\lambda=0$ to the corresponding classical operations.
\end{proof}

To answer the second question and show that the requirements are indeed not vacuous, we show that, quite
expectedly, the classical Dirac operator $\slashed{\mathrm{D}}_{(0)}$ fails to meet our criteria.
It will nevertheless be instructive to see which conditions are violated.
Furthermore, we will see that the deformed Dirac operator $\slashed{\mathrm{D}}^{\mathcal{F}}$ is ruled out
as well.
Let us f\/irst note that property~3), controlling the classical limit, is satisf\/ied by both of these
operators.
In order to understand if the classical Dirac operator satisf\/ies property~1), we expand up to f\/irst
order in the deformation parameter~$\lambda$, which yields
\begin{gather*}
\begin{split}
& \slashed{\mathrm{D}}_{(0)}(\psi\star a)=\slashed{\mathrm{D}}_{(0)}\left(\psi a+\frac{i\lambda}{2}
\Theta^{\alpha\beta}X_\alpha(\psi)X_\beta(a)\right)+\mathcal{O}\big(\lambda^2\big)
\\
& \phantom{\slashed{\mathrm{D}}_{(0)}(\psi\star a)}{}
=\slashed{\mathrm{D}}_{(0)}(\psi)a+\frac{i\lambda}{2}\Theta^{\alpha\beta}\slashed{\mathrm{D}}
_{(0)}\big(X_\alpha(\psi)\big)X_\beta(a)
\\
& \phantom{\slashed{\mathrm{D}}_{(0)}(\psi\star a)=}{}
{}+\iota_{V_{(0)}^{-1}\psi}(\mathrm{d}a)+\frac{i\lambda}{2}\Theta^{\alpha\beta}\iota_{V_{(0)}^{-1}
X_\alpha(\psi)}\big(X_\beta(\mathrm{d}a)\big)+\mathcal{O}\big(\lambda^2\big).
\end{split}
\end{gather*}
For $\slashed{\mathrm{D}}_{(0)}$ to be a~f\/irst-order noncommutative dif\/ferential operator the
$X_\alpha$ have to commute with~$\slashed{\mathrm{D}}_{(0)}$ and~$V_{(0)}$, i.e.\ the twist has to be
generated completely by Killing vector f\/ields.
This shows that for generic noncommutative Cartan geometries $(V,\Omega)$ on twisted manifolds
$(M,\mathcal{F})$ the classical Dirac operator $\slashed{\mathrm{D}}_{(0)}$ is {\it not} a~f\/irst-order
noncommutative dif\/ferential operator and in particular not a~noncommutative Dirac operator.

The deformed Dirac operator, on the other hand, is a~f\/irst-order noncommutative dif\/ferential operator,
since
\begin{gather*}
\slashed{\mathrm{D}}^{\mathcal{F}}\big(\psi\star a\big)=\slashed{\mathrm{D}}^{\mathcal{F}}
(\psi)\star a+\iota^\star_{iV_{(0)}^{-1}\star\psi}\big(\mathrm{d}a\big).
\end{gather*}
However, like the classical Dirac operator it fails to satisfy property 2), as in the construction of~$\slashed{\mathrm{D}}_{(0)}$ and~$\slashed{\mathrm{D}}^\mathcal{F}$ undeformed covariant dif\/ferentials
and contraction operators appear.
Summing up, our axioms above are satisf\/ied by $\slashed{\mathrm{D}}^\mathrm{AC}$ and
$\slashed{\mathrm{D}}^\mathrm{contr}$, but the deformed and classical Dirac operator fails in the general
to be a~noncommutative Dirac operator.
We will turn to the question for the remaining freedom to choose a~Dirac operator in the next sections.

\section{Comparing the noncommutative Dirac operators}\label{sec:non-uniqueness}

\subsection{Non-uniqueness in the general case}

Having established two examples of noncommutative Dirac operators on noncommutative Cartan geometries
$(V,\Omega)$ over twisted manifolds $(M,\mathcal{F})$, we shall now show that they do not coincide in general.
Our strategy is to calculate explicitly the two noncommutative Dirac operators
$\slashed{\mathrm{D}}^\mathrm{AC}$ and $\slashed{\mathrm{D}}^\mathrm{contr}$ for a~simple example of
$(V,\Omega)$ and $(M,\mathcal{F})$, from which the desired result can be directly read of\/f.

We will start with the $2$-dimensional case and consider the noncommutative spacetime known as `quantum plane'.
Let $M=\mathbb{R}^2$ and consider the twist $\mathcal{F}$ in~\eqref{eqn:twist},
constructed from $X_1 = t\partial_t$ and $X_2 = x\partial_x$, where $t$ and $x$ are global coordinates.
Notice that this twisted manifold $(M,\mathcal{F})$ leads to the commutation relations of the quantum
plane, i.e.\ $t\star x = e^{i\lambda} x\star t$.
We equip this twisted manifold with the following noncommutative Cartan geometry: $V=V^a\gamma_a =
\gamma_0\mathrm{d}t + \gamma_1 \mathrm{d}x$ and $\Omega =0$ is the unique $\star$-torsion free connection.
The $\star$-inverse $E_a$ of $V^a$ is def\/ined by $\iota_{E_a}^\star (V^b) = \delta_a^b$ and it is given
by $E_0 = \partial_t$, $E_1 = \partial_x$.
We further f\/ind for the $\star$-covariant dif\/ferential
$\mathrm{d}_\Omega \psi = \mathrm{d}\psi = \mathrm{d}t
\star e^{-\frac{i\lambda}{2} x\partial_x}\partial_t\psi + \mathrm{d}x\star
e^{\frac{i\lambda}{2}t\partial_t}\partial_x\psi $.
This leads to the following contraction Dirac operator~\eqref{eqn:2dcontrDirac}
\begin{gather}
\label{eqn:qpcontr}
\slashed{\mathrm{D}}^\mathrm{contr}\psi=i\left(\gamma^0e^{-\frac{i\lambda}{2}x\partial_x}
\partial_t\psi+\gamma^1e^{\frac{i\lambda}{2}t\partial_t}\partial_x\psi\right).
\end{gather}
In order to compare our two noncommutative Dirac operators, we also evaluate the Aschieri--Castellani Dirac
operator~\eqref{eqn:2dACDirac} for this model.
Using that $\mathrm{d}t\wedge_\star \mathrm{d}x = e^{\frac{i\lambda}{2}}\mathrm{d}t\wedge \mathrm{d}x = e^{\frac{i\lambda}{2}}\mathrm{vol}
/2$, with $\mathrm{vol} = \epsilon_{ab}V^a\wedge V^b$ denoting the volume form, we obtain for the inner
product~\eqref{eqn:2dinnerproduct}
\begin{gather}
\label{eqn:innerproductquantplane}
\ip{\psi_1}{\psi_2}=\cos(\lambda/2)\int_M\overline{\psi_1}\star\mathrm{vol}\star\psi_2.
\end{gather}
Furthermore, evaluating~\eqref{eqn:2dACDirac} we obtain
\begin{gather*}
\ip{\psi_1}{\slashed{\mathrm{D}}^\mathrm{AC}\psi_2}=i\int_M\overline{\psi_1}\star\mathrm{vol}
\star\left(e^{-\frac{i\lambda}{2}}\gamma^0e^{-\frac{i\lambda}{2}x\partial_x}
\partial_t\psi_2+e^{\frac{i\lambda}{2}}\gamma^1e^{\frac{i\lambda}{2}t\partial_t}\partial_x\psi_2\right),
\end{gather*}
which yields the Aschieri--Castellani Dirac operator on the quantum plane
\begin{gather}
\label{eqn:qpac}
\slashed{\mathrm{D}}^\mathrm{AC}\psi=\frac{i}{\cos(\lambda/2)}\Big(e^{-\frac{i\lambda}{2}}
\gamma^0e^{-\frac{i\lambda}{2}x\partial_x}\partial_t\psi+e^{\frac{i\lambda}{2}}
\gamma^1e^{\frac{i\lambda}{2}t\partial_t}\partial_x\psi\Big).
\end{gather}
Comparing~\eqref{eqn:qpcontr} and~\eqref{eqn:qpac} we observe that the noncommutative Dirac operators
$\slashed{\mathrm{D}}^\mathrm{contr}$ and $\slashed{\mathrm{D}}^\mathrm{AC}$ do not coincide.
Notice that the dif\/ference is not just in the overall factor, but the two terms have also acquired
dif\/ferent phases.
With the $4$-dimensional analog of the quantum plane and a~similar calculation, we obtain the following
result.
\begin{propo}
The two noncommutative Dirac operators $\slashed{\mathrm{D}}^\mathrm{AC}$ and
$\slashed{\mathrm{D}}^{\mathrm{contr}}$ do not coincide for generic noncommutative Cartan geometries
$(V,\Omega)$ over twisted manifolds $(M,\mathcal{F})$ of dimen\-sion~$2$ or~$4$.
\end{propo}

\subsection{The formal self-adjointness condition}

In regard of the non-uniqueness result above we try to include stronger conditions on noncommutative Dirac
operators in order to single out a~preferred choice.
We shall focus in this subsection only on one strongly motivated extra condition, which is formal
self-adjointness.
This condition is essential for associating to a~noncommutative Dirac operator a~quantum f\/ield theory of
noncommutative Dirac f\/ields, see Section~\ref{sec:quantization}.
Notice that also in a~Riemannian setting, a~formal self-adjointness condition is an important ingredient
for understanding the spectral theory of Dirac operators.
\begin{defi}
Let $(M,\mathcal{F})$ be a~twisted manifold and $(V,\Omega)$ a~noncommutative Cartan geo\-met\-ry on
$(M,\mathcal{F})$.
A noncommutative Dirac operator $\slashed{\mathrm{D}}$ is called {\it formally self-adjoint}, if it
satisf\/ies $\ip{\psi_1}{\slashed{\mathrm{D}}\psi_2} = \ip{\slashed{\mathrm{D}}\psi_1}{\psi_2}$, for all
Dirac f\/ields $\psi_1$, $\psi_2$ with compactly overlapping support.
Here~$\ip{\cdot}{\cdot}$ is the inner product def\/ined for $D=2$ in~\eqref{eqn:2dinnerproduct} and
for $D=4$ in~\eqref{eqn:innerproduct}.
\end{defi}

Notice that the classical Dirac operator $\slashed{\mathrm{D}}_{(0)}$ is formally self-adjoint with respect
to the classical inner product $\ip{\cdot}{\cdot}\vert_{\lambda=0}^{}$ if the spin connection is
torsion free.
For the Aschieri--Castellani Dirac operator we obtain the analogous property in the twisted setting.
\begin{propo}
\label{prop:ACformalSA}
Let $(M,\mathcal{F})$ be any twisted manifold of dimension $2$ or $4$, and let $(V,\Omega)$ be
a~noncommutative Cartan geometry that is $\star$-torsion free, i.e.\ $\mathrm{d}_\Omega V = 0$.
Then the noncommutative Dirac operator defined in~\eqref{eqn:2dACDirac} and~\eqref{eqn:ACDirac} is
formally self-adjoint.
\end{propo}
\begin{proof}
We show this statement by the following calculation for the $2$-dimensional case
\begin{gather*}
\ip{\slashed{\mathrm{D}}^\mathrm{AC}\psi_1}{\psi_2}
=\ip{\psi_2}{\slashed{\mathrm{D}}^\mathrm{AC}\psi_1}^\ast=2i\int_{M}\mathrm{Tr}
\big(\gamma_3^\dagger V^\dagger\star\gamma_0\psi_2\wedge_\star\mathrm{d}_\Omega\overline{\psi_1}\gamma_0\big)
\\
\phantom{\ip{\slashed{\mathrm{D}}^\mathrm{AC}\psi_1}{\psi_2}}{}
=2i\int_{M}\mathrm{Tr}\left(\gamma_3\gamma_0V\star\psi_2\wedge_\star\mathrm{d}
_\Omega\overline{\psi_1}\gamma_0\right)=-2i\int_M\mathrm{Tr}\left(\psi_2\star\mathrm{d}
_\Omega\overline{\psi_1}\wedge_\star V\gamma_3\right)
\\
\phantom{\ip{\slashed{\mathrm{D}}^\mathrm{AC}\psi_1}{\psi_2}}{}
=-2i\int_M\mathrm{d}\mathrm{Tr}\left(\psi_2\star\overline{\psi_1}
\wedge_\star V\gamma_3\right)+2i\int_M\mathrm{Tr}\left(\mathrm{d}_\Omega\psi_2\star\overline{\psi_2}
\wedge_\star V\gamma_3\right)
\\
\phantom{\ip{\slashed{\mathrm{D}}^\mathrm{AC}\psi_1}{\psi_2}}{}
=\ip{\psi_1}{\slashed{\mathrm{D}}^\mathrm{AC}\psi_2}.
\end{gather*}
In the f\/irst equality we have used hermiticity of the inner product.
In the second equality we have used that $\ast$ is a~graded involution on the deformed dif\/ferential forms
$(\Omega^\bullet(M),\wedge_\star,\mathrm{d})$ as well as $(\mathrm{d}_\Omega\psi_{1}\star
\overline{\psi_2})^\dagger = \gamma_0\psi_2\star \mathrm{d}_\Omega\overline{\psi_1}\gamma_0$, which
follows from the hermiticity condition $\Omega^\dagger = -\gamma_0\Omega\gamma_0$.
Then for the third step we used $V^\dagger = \gamma_0V\gamma_0$, $\gamma_3^\dagger=\gamma_3$ and
$\gamma_0^2=1$, and in the fourth one graded cyclicity~\eqref{eqn:gradedcyc} and twice $\gamma_3\gamma_a =
- \gamma_a\gamma_3$.
In the f\/ifth equality we made use of the graded Leibniz rule of $\mathrm{d}_\Omega$ and the
$\star$-torsion constraint $\mathrm{d}_\Omega V=0$.
The last step is simply Stokes' theorem.
The proof for the $4$-dimensional case is once again fully analogous.
\end{proof}

We will now show that our second example of a~noncommutative Dirac operator, that is the contraction Dirac
operator, does not satisfy the formal self-adjointness condition on generic $\star$-torsion free
noncommutative Cartan geometries $(V,\Omega)$.
To this end we again consider the quantum plane as a~simple example.
The contraction Dirac operator is given in $\eqref{eqn:qpcontr}$ and using the explicit form of the inner
product~\eqref{eqn:innerproductquantplane} we obtain for the formal adjoint of the contraction Dirac
operator
\begin{gather*}
(\slashed{\mathrm{D}}^{\mathrm{contr}})^{\ast}\psi
=i\left(e^{-i\lambda}\gamma^0e^{-\frac{i\lambda}{2}x\partial_x}\partial_t\psi
+e^{i\lambda}\gamma^1e^{\frac{i\lambda}{2}t\partial_t}\partial_x\psi\right).
\end{gather*}
The dif\/ferential operators $(\slashed{\mathrm{D}}^{\mathrm{contr}})^{\ast} $ and
$\slashed{\mathrm{D}}^{\mathrm{contr}}$ (cf.~\eqref{eqn:qpcontr}) do not agree, hence the contraction Dirac
operator is not formally self-adjoint on the quantum plane.
With the analogous investigation of the $4$-dimensional analog of the quantum plane, this leads to the
following conclusion.
\begin{propo}
The noncommutative Dirac operator $\slashed{\mathrm{D}}^\mathrm{contr}$ is not formally self-adjoint for
generic $\star$-torsion free noncommutative Cartan geometries $(V,\Omega)$ over twisted manifolds
$(M,\mathcal{F})$ of dimension $2$ or $4$.
Hence, the Aschieri--Castellani Dirac operator is the preferred choice among our two noncommutative Dirac
operators.
\end{propo}

\subsection{Semi-Killing deformations as a~special case}

As noted in~\cite{Aschieri:2009qh,Ohl:2009pv, Schupp:2009pt}, any metric f\/ield solving the classical
Einstein equations also solves the noncommutative Einstein equations~\cite{Aschieri:2005zs} if the twist is
semi-Killing.
Explicitly, an Abelian twist~\eqref{eqn:twist} is semi-Killing if $\Theta^{\alpha\beta} X_\alpha\otimes
X_\beta \in \Xi \otimes \mathfrak{K} + \mathfrak{K} \otimes \Xi$, where $\Xi$ is the Lie algebra of vector
f\/ields on $M$ and $\mathfrak{K}:=\lbrace X\in\Xi: \mathcal{L}_X (V)=0 \ \text{and} \ \mathcal{L}_X(\Omega)=0\rbrace$ is the Killing Lie algebra.
Using the canonical form of $\Theta^{\alpha\beta}$~\eqref{eqn:canonical}, this condition is equivalent to
requiring that either~$X_{2n}$ or~$X_{2n-1}$ is a~Killing vector f\/ield, for all $n=1,2,\dots$.

Let now $(V,\Omega)$ be a~noncommutative Cartan geometry over a~twisted manifold $(M,\mathcal{F})$, such
that $\mathcal{F}$ is semi-Killing.
We notice that in this case $(V= V^a\gamma_a,\Omega = \frac{1}{4}\omega^{ab}\gamma_{ab})$ solves the
noncommutative Einstein equations (in vielbein form)~\cite{Aschieri:2009ky} whenever it solves the
commutative Einstein equations.
In the following we study our noncommutative Dirac operators for this particular class of examples and show
that they coincide.
Let us denote by $E_a$ the basis for the vector f\/ields on $M$ which is specif\/ied by the undeformed
contraction condition $\iota_{E_a}( V^b) =\delta_a^b$.
This condition implies that $\mathcal{L}_X (E_a)=0$ for all $X\in\mathfrak{K}$ and, hence, also the
deformed contraction condition $\iota^\star_{E_a} (V^b) =\delta_a^b$ holds true for semi-Killing twists.
The $\star$-inverse vierbein therefore reads $V^{-1_\star} = E_a \gamma^a$.
We def\/ine the components $\Omega_a$ of the spin connection by $\Omega =: V^a \Omega_a$.
The conditions $\mathcal{L}_X(\Omega)=0$ and $\mathcal{L}_X(V)=0$ imply that $\mathcal{L}_X (\Omega_a) =0$,
for all $X\in\mathfrak{K}$, and thus $\Omega = V^a\Omega_a =V^a\star\Omega_a$.
Furthermore, we def\/ine the dif\/ferential operator $E_a^\star$ by $E_a^\star (\psi):=
\iota^\star_{E_a}(\mathrm{d}\psi)$ and we obtain that
$\iota_{E_a}^\star(\mathrm{d}_\Omega\psi)=E^\star_a(\psi)-\Omega_a\star\psi$.
The contraction Dirac operator~\eqref{eqn:2dcontrDirac} expressed in this basis reads
\begin{gather}
\label{eqn:contrsemikilling}
\slashed{\mathrm{D}}^\mathrm{contr}\psi=i\gamma^a\iota_{E_a}^\star\big(\mathrm{d}
_\Omega\psi\big)=i\gamma^a\big(E_a^\star(\psi)-\Omega_a\star\psi\big).
\end{gather}
Let us also compute explicitly the Aschieri--Castellani Dirac operator~\eqref{eqn:2dACDirac} for this class
of examples.
We show the calculation for the $2$-dimensional case and note that it is fully analogous for $D=4$.
Due to the semi-Killing property we have $V\wedge_\star V\gamma_3=V\wedge V \gamma_3= \mathrm{vol}$, with
$\mathrm{vol}= \epsilon_{ab} V^a\wedge V^b$.
Hence, the inner product~\eqref{eqn:2dinnerproduct} reads
\begin{gather}
\label{eqn:tmpsemikillinginnerproduct}
\ip{\psi_1}{\psi_2}=\int_M\overline{\psi_1}\star\mathrm{vol}\star\psi_2.
\end{gather}
Using $V\gamma_3 = -V^a\epsilon_{ab}\gamma^b$ we f\/ind that~\eqref{eqn:2dACDirac} simplif\/ies to
\begin{gather*}
\ip{\psi_1}{\slashed{\mathrm{D}}^\mathrm{AC}\psi_2}=2i\int_M\overline{\psi_1}\star V^a\epsilon_{ab}
\gamma^b\wedge_\star\mathrm{d}_\Omega\psi_2=\int_M\overline{\psi_1}\star\mathrm{vol}
\star i\gamma^a\iota_{E_a}^\star\big(\mathrm{d}_\Omega\psi_2\big).
\end{gather*}
This and~\eqref{eqn:tmpsemikillinginnerproduct} implies that $\slashed{\mathrm{D}}^{\mathrm{AC}}$ coincides
with $\slashed{\mathrm{D}}^\mathrm{contr}$ (cf.~\eqref{eqn:contrsemikilling}) for semi-Killing twists.
Notably, also the deformed Dirac operator $\slashed{\mathrm{D}}^\mathcal{F}$ (obtained by deforming the
classical Dirac opera\-tor~$\slashed{\mathrm{D}}_{(0)}$ corresponding to $(V,\Omega)$) coincides with
$\slashed{\mathrm{D}}^{\mathrm{AC}}$ and $\slashed{\mathrm{D}}^\mathrm{contr}$, even though it in general
fails to satisfy our requirements.
We would like to stress that for generic semi-Killing twists the noncommutative Dirac operators do not
coincide with the classical one $\slashed{\mathrm{D}}_{(0)}$ corresponding to~$(V,\Omega)$,
cf.~Section~\ref{sec:examples} for explicit examples.
In summary, we have obtained the following
\begin{propo}
For semi-Killing twists the two noncommutative Dirac operators~$\slashed{\mathrm{D}}^{\mathrm{AC}}$ and $\slashed{\mathrm{D}}^\mathrm{contr}$ coincide.
Even more, in this case these operators also coincide with the operator~$\slashed{\mathrm{D}}^\mathcal{F}$,
which in general does not satisfying our axioms.
For practical purposes one can therefore choose the technically most convenient one.
\end{propo}

If, moreover, all vector f\/ields $X_\alpha$ in the twist $\mathcal{F}$ are Killing, then the
dif\/ferential operator $E_a^\star$ def\/ined by $E^\star_a(\psi) = \iota_{E_a}^\star(\mathrm{d} \psi) $
coincides with the vector f\/ield $E_a$ (the inverse vierbein).
Since also $\Omega_a\star \psi =\Omega_a\psi$, we obtain for the contraction Dirac
operator~\eqref{eqn:contrsemikilling} $\slashed{\mathrm{D}}^\mathrm{contr}\psi = i\gamma^a(E_a(\psi)
- \Omega_a\psi) $.
Hence, the operator $\slashed{\mathrm{D}}^\mathrm{contr}$ coincides in the Killing case with the classical
Dirac operator $\slashed{\mathrm{D}}_{(0)}$ corresponding to $(V,\Omega)$.
Since actual Killing twists are contained in the class of semi-Killing twists, we have further
$\slashed{\mathrm{D}}^\mathrm{contr} = \slashed{\mathrm{D}}^\mathrm{AC} = \slashed{\mathrm{D}}^\mathcal{F}
=\slashed{\mathrm{D}}_{(0)}$ for Killing twists.

Applying this result to the noncommutative (Riemannian) space considered in~\cite{Connes:2000tj}, which in
our notation corresponds to a~Killing twist deformation of the sphere, we also come to the conclusion that
the classical Dirac operator is a~suitable noncommutative Dirac operator for this model.
Moreover, all of our deformed constructions for noncommutative Dirac operators reduce in this case to the
classical one, hence there is (within our class of operators) no alternative choice.
The main reason behind this is, of course, the invariance of the Cartan geometry and the classical Dirac
operator under the (Killing) vector f\/ields entering the twist.

\section{Quantum f\/ield theory of noncommutative Dirac f\/ields}\label{sec:quantization}

In~\cite{Ohl:2009qe} we have constructed retarded/advanced Green's operators as well as the solution space
of deformed wave equations, which led to the construction of noncommutative Klein--Gordon quantum f\/ield
theories.
In this section we generalize these results to noncommutative Dirac operators, and hence to noncommutative
quantum f\/ield theories of Dirac f\/ields.

We start with a~general linear dif\/ferential operator $P:C^\infty(M,\mathbb{C}^N)\to
C^\infty(M,\mathbb{C}^N)$ acting on Dirac f\/ields $\psi \in C^\infty(M,\mathbb{C}^N)$, with the classical
limit $P_{(0)}:= P\vert_{\lambda=0}^{} = \slashed{\mathrm{D}}_{(0)} + m$, where $m\in\mathbb{R}$ is a~mass
term.
Examples are noncommutative Dirac operators in the sense of Def\/inition~\ref{defi:Dirac} with an
additional mass term, i.e.\ $P=\slashed{\mathrm{D}} + m$.
It is well known that for the classical massive Dirac operator $P_{(0)}=\slashed{\mathrm{D}}_{(0)}+m:
C^\infty(M,\mathbb{C}^N)\to C^\infty(M,\mathbb{C}^N)$ on globally hyperbolic spacetimes there exists
a~unique retarded and advanced Green's operator $G_{(0)}^\pm$, see e.g.~\cite{Muehlhoff:2010ra} for a~proof
employing a~modern language.
We remind the reader that a~retarded/advanced Green's operator is a~linear map $G_{(0)}^\pm:
C_0^\infty(M,\mathbb{C}^N)\to C^\infty(M,\mathbb{C}^N)$ on compactly supported functions, which satisf\/ies
the inhomogeneous equation of motion $G_{(0)}^\pm \circ P_{(0)} = P_{(0)}\circ G_{(0)}^\pm =\mathrm{id}$
and the support condition $\mathrm{supp}(G_{(0)}^\pm \varphi) \subseteq J^\pm( \mathrm{supp}(\varphi))$,
for all $\varphi\in C_0^\infty(M,\mathbb{C}^N)$, where $J^\pm(\mathrm{supp}(\varphi))$ is the
forward/backward lightcone of the set $\mathrm{supp}(\varphi)$.
If we assume that our noncommutative Cartan geometry $(V,\Omega)$ over our twisted manifold
$(M,\mathcal{F})$ is such that $V_{(0)} = V\vert_{\lambda=0}^{}$ is the vielbein of a~globally hyperbolic
Lorentzian metric on $M$, then we can f\/ind a~unique retarded and advanced Green's operator $G^\pm$ for
$P$.
Let us write $P = \sum\limits_{n=0}^\infty \lambda^n P_{(n)}$, then a~construction as in~\cite[Theorem~1]{Ohl:2009qe} shows that the Green's operators $G^\pm = \sum\limits_{n=0}^\infty \lambda^n G_{(n)}^\pm$
for $P$ are given by, for~$n\geq 1$,
\begin{gather}
\label{eqn:greens}
G^\pm_{(n)}:=\sum\limits_{k=1}^n\sum\limits_{j_1=1}^n\cdots\sum\limits_{j_k=1}
^n(-1)^k\delta_{j_1+\cdots+j_k,n}G_{(0)}^\pm\circ P_{(j_1)}\circ\cdots\circ G_{(0)}^\pm\circ P_{(j_k)}
\circ G_{(0)}^\pm,
\end{gather}
where $\delta_{n,m}$ is the Kronecker delta.
In an earlier version of this manuscript, as well as in~\cite[Theorem~1]{Ohl:2009qe}, it was assumed that
$P_{(n)}$, $n\geq 1$, are dif\/ferential operators of {\it compact} support.
However, due to recent advances in understanding Green-hyperbolic operators~\cite{Baernew}, this condition
turns out to be unnecessary, since the compositions in~\eqref{eqn:greens} are well-def\/ined also for
generic dif\/ferential operators $P_{(n)}$.

Using the Green's operators for $P$, we can characterize the solution space $\mathrm{Sol}:= \{\psi \in
C_\mathrm{sc}^\infty(M,\mathbb{C}^N): P \psi =0\}$, where the subscript $_\mathrm{sc}$ denotes
functions of spacelike compact support.
All solutions are obtained by the causal propagator $G:= G^+ - G^-:C^\infty_0(M,\mathbb{C}^N)\to
C^\infty_\mathrm{sc}(M,\mathbb{C}^N)$, since the following sequence of linear maps is an exact complex
\begin{gather}
\label{eqn:sequence}
\xymatrix{\{0\}\ar[r]&C^\infty_0\big(M,\mathbb{C}^N\big)\ar[r]^-{P}&C^\infty_0\big(M,\mathbb{C}^N\big)\ar[r]^-{G}
&C^\infty_{\mathrm{sc}}\big(M,\mathbb{C}^N\big)\ar[r]^-{P}&C^\infty_\mathrm{sc}\big(M,\mathbb{C}^N\big).
}
\end{gather}
The proof of this statement is similar to the one of~\cite[Theorem 2]{Ohl:2009qe}, hence we can omit it
here.

With these tools we can construct the canonical anti-commutation relation (CAR) algebra corresponding to
our dif\/ferential operators $P: C^\infty(M,\mathbb{C}^N) \to C^\infty(M,\mathbb{C}^N) $.
This is the observable algebra of the quantized noncommutative Dirac f\/ield.
For this construction we also require a~hermitian inner product $\ip{\cdot}{\cdot}$ on
$C^\infty(M,\mathbb{C}^N)$ and that $P$ is formally self-adjoint.
In $D=2$ and $D=4$ the inner product is given in~\eqref{eqn:2dinnerproduct} and~\eqref{eqn:innerproduct},
respectively.
By Proposition~\ref{prop:ACformalSA} the Aschieri--Castellani Dirac operator is formally self-adjoint in
$D=2$ and $D=4$ for $\star$-torsion free $(V,\Omega)$.
The same holds true for the massive Aschieri--Castellani Dirac operator $\slashed{\mathrm{D}}^{\mathrm{AC}}
+m$, which thus provides an example for a~dif\/ferential operator $P$ with the properties we are looking
for.

For constructing the CAR algebra, let us def\/ine another inner product on $C^\infty_0(M,\mathbb{C}^N)$ by
using the causal propagator $G = G^+ - G^-$ corresponding to $P$, for all $\varphi_1,\varphi_2\in
C^\infty_0(M,\mathbb{C}^N) $,
\begin{gather*}
\langle\langle\varphi_1,\varphi_2\rangle\rangle:=i\ip{\varphi_1}{G\varphi_2}.
\end{gather*}
Since $\ip{\varphi_1}{\varphi_2}^\ast = \ip{\varphi_2}{\varphi_1}$ and $P$ is by assumption formally
self-adjoint (which implies that $G$ is formally skew-adjoint) we obtain
\begin{gather*}
\langle\langle\varphi_1,\varphi_2\rangle\rangle^\ast=-i\ip{G\varphi_2}{\varphi_1}=i\ip{\varphi_2}
{G\varphi_1}=\langle\langle\varphi_2,\varphi_1\rangle\rangle.
\end{gather*}
Due to the exact sequence~\eqref{eqn:sequence} and the fact that $G$ is formally skew-adjoint,
$\langle\langle\cdot,\cdot\rangle\rangle$ induces a~hermitian inner product on the quotient $H:=
C^\infty_0(M,\mathbb{C}^N)/ P[C^\infty_0(M,\mathbb{C}^N)]$.
Furthermore, the inner product $\langle\langle \cdot,\cdot\rangle\rangle$ on $H$ is
positive-def\/inite.
To show this statement let us consider the classical limit $\lambda=0$.
We f\/ind by using Green's formula~\cite[p.~160, Proposition~9.1]{Taylor} that the inner product in this limit
is, for all $\varphi_1,\varphi_2\in H\vert_{\lambda=0}= C^\infty_0(M,\mathbb{C}^N)/
P_{(0)}[C^\infty_0(M,\mathbb{C}^N)] $,
\begin{gather*}
\nonumber\langle\langle\varphi_1,\varphi_2\rangle\rangle\vert_{\lambda=0}
=i\ip{P_{(0)}G_{(0)}^\pm\varphi_1}{G_{(0)}\varphi_2}\vert_{\lambda=0}
=i\int_\Sigma\overline{i\gamma_a n^a G_{(0)}\varphi_1}G_{(0)}\varphi_2\mathrm{vol}_\Sigma
\\
\phantom{\nonumber\langle\langle\varphi_1,\varphi_2\rangle\rangle\vert_{\lambda=0}}{}
=\int_\Sigma(G_{(0)}\varphi_1)^\dagger G_{(0)}\varphi_2\mathrm{vol}_\Sigma.
\end{gather*}
Here $\Sigma$ is any Cauchy surface and $n = E_an^a$ its future-pointing normal vector f\/ield.
In the last equality we have used that we can choose $n=E_0$.
Since the inner product $\langle\langle\cdot,\cdot \rangle\rangle$ is positive-def\/inite at order
$\lambda^0$ it is positive-def\/inite to all orders in the deformation parameter.

The inner-product space $(H,\langle\langle\cdot,\cdot\rangle\rangle)$ can be quantized in
terms of a~CAR algebra, see e.g.~\cite{Bar:2011iu} for a~modern review of these techniques: To any element
$\varphi\in H$ we associate an abstract operator~$a(\varphi)$ and consider the free unital $\ast$-algebra
$A^\mathrm{free}$ generated by all~$a(\varphi)$,~$\varphi\in H$.
We def\/ine the CAR algebra $A^{\text{CAR}}:= A^\mathrm{free}/\mathcal{I}$ as the quotient of
$A^\mathrm{free}$ by the both-sided $\ast$-ideal $\mathcal{I}$ generated by the elements, for all
$\varphi_1,\varphi_2\in H$ and $\alpha_1,\alpha_2\in \mathbb{C}$,
\begin{subequations}
\begin{gather}
\label{1}
a\big(\alpha_1\varphi_1+\alpha_2\varphi_2\big)-\alpha_1a(\varphi_1)-\alpha_2a(\varphi_2),
\\
\label{2}
\big\{a(\varphi_1),a(\varphi_2)\big\},
\\
\label{3}
\big\{a(\varphi_1)^\ast,a(\varphi_2)\big\}-\langle\langle\varphi_1,\varphi_2\rangle\rangle1,
\end{gather}
\end{subequations}
where $\{\cdot,\cdot\}$ is the anti-commutator.
The interpretation of this quotient is as follows:~\eqref{1} allows us to regard $a(\psi)$ as smeared
linear f\/ield operators.
\eqref{2} and~\eqref{3} encode the CAR.
The on-shell condition is already implemented in $H$.
In the physics literature, the Dirac f\/ield operator is typically denoted by $\Psi(x)$ and its adjoint by
$\overline{\Psi}(x)$.
This notation is related to ours by $a(\varphi) = \ip{\Psi}{\varphi}$ and $a(\varphi)^\ast =
\ip{\varphi}{\Psi}$, where by the inner products we (formally) denote the smearing of the f\/ield operators
by test functions.
In this notation~\eqref{2} states that $\Psi(x)$ anti-commutes with $\Psi(y)$ and~\eqref{3} that the
anti-commutator between $\overline{\Psi}(x)$ and $\Psi(y)$ is non-trivial.

We end this section with a~comparison to the more conventional approach to quantum f\/ield theory on
noncommutative spacetimes as followed e.g.\
in~\cite{Aschieri:2007sq,Balachandran:2007vx,Fiore:2007vg, Zahn:2006wt}, where the algebra of f\/ield
operators is also deformed by the twist.
This is motivated by the desire for a~representation of the Moyal--Weyl deformed Poincar{\'e} Hopf algebra
on this algebra.
While for highly symmetric models like the Moyal--Weyl Minkowski spacetime this is certainly an interesting
approach, it unfortunately does not generalize to situations where there are no isometries, or where the
twist is generated not by Killing vector f\/ields alone.
The relevant argument, which has been given already in the appendix of~\cite{Schenkel:2010jr}, is that
generic vector f\/ields on spacetime can not be represented on the algebra of f\/ield operators.
Our formulation of noncommutative quantum f\/ield theory is complementary to this approach: It is valid for
generic twisted curved spacetimes, but it obscures the role of twisted symmetry Hopf algebras in the very
special cases where they are available.

\section{Explicit examples}\label{sec:examples}

In this section we will explicitly study the noncommutative Dirac operators discussed in
Section~\ref{sec:definitionDirac} on two noncommutative (curved) spacetimes.
For their attractive features, e.g.\ as solutions to noncommutative Einstein equations, we will focus on
semi-Killing deformations.
As shown in Section~\ref{sec:non-uniqueness}, our examples of noncommutative Dirac operators
coincide in this case and we will collectively denote them by $\slashed{\mathrm{D}}$.
These studies are complementing our explicit examples of deformed Klein--Gordon
operators~\cite{Schenkel:2010sc}.

\subsection[$\kappa$-Minkowski spacetime]{$\boldsymbol{\kappa}$-Minkowski spacetime}

As a~f\/irst example we consider $M=\mathbb{R}^4$ with global coordinates denoted by $x^\mu=(t,x^j)$ and
the Minkowski vierbein $V = \gamma_a\delta^a_\mu \mathrm{d}x^\mu$, along with the spin connection $\Omega =0$.
For the twist~\eqref{eqn:twist} we use $X_1 = \partial_t$ and $X_2 = x^j\partial_j$, which yields
a~semi-Killing twist.
The commutation relations of the coordinate functions are those of $\kappa$-Minkowski spacetime,
i.e.\ $\starcom{t}{x^j} = i\lambda x^j$ and $\starcom{x^i}{x^j}=0$.
This example has been studied intensively in the literature, see
e.g.~\cite{Borowiec:2008uj,Bu:2006dm,Govindarajan:2008qa,Juric:2012xt,Kim:2008mp,Meljanac:2012fa}.
Furthermore, various f\/ields with their equation of motion operators have been studied on this particular
noncommutative spacetime, see~\cite{Schenkel:2010sc} for the scalar f\/ield and~\cite{Dimitrijevic:2011jg}
for the ${\rm U}(1)$ gauge f\/ield.
We supplement these studies by the Dirac f\/ield with equation of motion operator given by any of the
noncommutative Dirac operators introduced in Section~\ref{sec:definitionDirac}, which all coincide
for this model since $\mathcal{F}$ is semi-Killing.
For spectral triple approaches to Dirac operators on the $\kappa$-Minkowski space we refer the reader
to~\cite{spectral1,spectral2}.

Using that $\mathcal{L}_{X_1}(V^a) =0$, $\mathcal{L}_{X_2}(V^0) =0$ and $\mathcal{L}_{X_2}(V^j) = V^j$ we
obtain for the $\star$-contraction the following expression
\begin{gather*}
\iota^\star_{E_0}(\mathrm{d}_\Omega\psi)=\partial_t\psi,\qquad \iota_{E_j}^\star(\mathrm{d}_\Omega\psi)
=e^{\frac{i\lambda}{2}\partial_t}\partial_j\psi.
\end{gather*}
Since our noncommutative Dirac operators coincide for this model we choose to calculate the simplest one,
which is the contraction Dirac operator~\eqref{eqn:2dcontrDirac}, and f\/ind
\begin{gather*}
\slashed{\mathrm{D}}\psi
=i\gamma^a\iota_{E_a}^\star(\mathrm{d}_\Omega\psi)
=i\left(\gamma^0\partial_t\psi+\gamma^j e^{\frac{i\lambda}{2}\partial_t}\partial_j\psi\right).
\end{gather*}
For the solutions of the noncommutative Dirac equation $ \slashed{\mathrm{D}}\psi =0$ we can derive
a~dispersion relation by squaring the equation of motion operator $\slashed{\mathrm{D}}$.
More explicitly, this yields
\begin{gather*}
\square:=-\slashed{\mathrm{D}}^2=\partial_t^2-\bigtriangleup e^{i\lambda\partial_t},
\end{gather*}
where $\bigtriangleup:= \partial_1^2 + \partial_2^2 +\partial_3^2$ is the spatial Laplacian.
To study the dispersion relation we make a~plane wave ansatz $\psi = \chi e^{i (E t + k_j x^j)}$, where
$E$ is the energy, $k_j$ the momentum and $\chi\in \mathbb{C}^4$ a~polarization spinor.
Since $\slashed{\mathrm{D}} \psi=0$ implies $\square\psi =0$ we obtain the deformed energy-momentum relation
\begin{subequations}
\begin{gather}
\label{eqn:k-Minkowski-dispersion}
E^2-e^{-\lambda E}{\bf k}^2=0
\qquad
\Leftrightarrow
\qquad
E^2e^{\lambda E}={\bf k}^2.
\end{gather}
From the equation of motion $\slashed{\mathrm{D}} \psi=0$ we further obtain a~condition on the polarization
spinor
\begin{gather}
\label{eqn:k-Minkowski-spin}
\left(\gamma^0E+\gamma^j k_je^{-\frac{\lambda}{2}E}\right)\chi=0.
\end{gather}
\end{subequations}
Without loss of generality we choose the spatial momentum along the third direction, i.e.\ $\mathbf{k} =
(0,0,k)$, such that~\eqref{eqn:k-Minkowski-spin} becomes $(\gamma^0E + \gamma^3 k
e^{-\frac{\lambda}{2} E} )\chi =0$.
Using the on-shell condition~\eqref{eqn:k-Minkowski-dispersion} this becomes independent of $\lambda$ and
reduces to the analogous equation in the commutative case.
We thus f\/ind that the physical spin polarizations, which are characterized as the solutions
of~\eqref{eqn:k-Minkowski-spin}, do not receive noncommutative corrections.
Hence, this type of noncommutative geometry does not introduce an anomalous spin precession.

\subsection{Noncommutative anti de Sitter space}

We now turn to a~curved spacetime example, for which the natural f\/irst candidates are the maximally
symmetric (anti) de Sitter ((A)dS) spacetimes.
We choose AdS which is of relevance for model building in particle physics and AdS/CFT, but note that
similar calculations for the cosmologically relevant dS are fully analogous.
A particle-physics model employing a~deformation of AdS can be found in~\cite{Ohl:2010bh}.

We focus on the Poincar\'{e} patch of 4-dimensional AdS, that is, $M = \mathbb{R}^3 \times (0,\infty)$ with
coordinates $x^\mu = (x^i,z)$ and the vierbein $V = \gamma_aR z^{-1}\delta^a_\mu \mathrm{d}x^\mu$.
The generalization to higher dimensions is straightforward and just amounts to using the higher dimensional
Clif\/ford algebras.
In the following we f\/ix the radius of curvature to $R=1$, and for the gamma-matrices we denote the
contraction with the (inverse) vielbein explicitly by a~hat, e.g.\ $\hat\gamma^\mu:= E_a^\mu \gamma^a =
z\delta_a^\mu \gamma^a$.
To deform this space we employ the twist~\eqref{eqn:twist} with the $2N$ mutually commuting vector f\/ields
$X_\alpha$, $\alpha=1,\dots,2N$, given by
\begin{gather}
\label{eqn:rsvectors}
X_{2n-1}=T_{2n-1}^{i}\partial_i,
\qquad
X_{2n}=\vartheta(z)T_{2n}^{i}\partial_i,
\qquad
n=1,2,\dots,N.
\end{gather}
In this expression the $T_\alpha^i$ are real numbers and $\vartheta(z)\in C^\infty(0,\infty)$ is a~real
valued function.
Notice that this twist is semi-Killing, since all $X_{2n-1}$ are Killing vector f\/ields.
The $\star$-commutation relations of the coordinate functions $(x^i,z)$ read
\begin{gather}
\label{eqn:comads}
\starcom{x^i}{x^j}=i\lambda\vartheta(z)\Theta^{\alpha\beta}T_\alpha^iT_\beta^j
,
\qquad
\starcom{x^i}{z}=0.
\end{gather}
Hence, this model describes a~$z$-dependent Moyal--Weyl deformation of the $\mathbb{R}^3$ hypersurfaces at
constant $z$.
The $\star$-torsion free spin connection is $\Omega = -\frac{1}{2}V^i \gamma_{i3}$.

To compute the $\star$-contraction $\iota_{E_a}^\star(\mathrm{d}_\Omega \psi)$ we f\/irst notice that
$\mathcal{L}_{X_{2n-1}}(V^a) =0$, $\mathcal{L}_{X_{2n}}(V^i) =\vartheta^\prime(z)T_{2n}^iV^3$ and
$\mathcal{L}_{X_{2n}}(V^3) =0$.
We then obtain
\begin{gather*}
\iota_{E_i}^\star(\mathrm{d}_\Omega\psi)
=z\partial_i\psi+\frac{\gamma_{i3}}{2}\psi,\qquad \iota_{E_3}^\star(\mathrm{d}_\Omega\psi)
=z\partial_z\psi+\frac{i\lambda}{2}z\vartheta^\prime(z)\mathcal{T}\psi,
\end{gather*}
where $\mathcal{T}:= \mathcal{T}^{ij}\partial_i\partial_j:= \sum\limits_{n=1}^N
T_{2n}^iT_{2n-1}^j\partial_i\partial_j$.
Since the twist is semi-Killing and our examples of noncommutative Dirac operators are therefore
equivalent, we once again choose~\eqref{eqn:2dcontrDirac} as the technically most convenient one, and f\/ind
\begin{gather}
\label{eqn:diracads}
\slashed{\mathrm{D}}\psi=i\gamma^a\iota_{E_a}^\star(\mathrm{d}_\Omega\psi)=\slashed{\mathrm{D}}_{(0)}
\psi+\frac{\lambda}{2}z\vartheta^\prime(z)\gamma_3\mathcal{T}\psi,
\end{gather}
where $ \slashed{\mathrm{D}}_{(0)} $ is the classical Dirac operator.
This is the equation of motion operator stemming from the action
\begin{gather}
\label{eqn:actionads}
S=\ip{\psi}{\slashed{\mathrm{D}}\psi}=\int_M\overline{\psi}\slashed{\mathrm{D}}\psi\mathrm{vol},
\end{gather}
where in the last equality we have used that for the present model the inner
product~\eqref{eqn:innerproduct} coincides with the undeformed one.

A crucial point for the construction of quantum f\/ields on AdS is the existence of a~f\/inite inner
product, which is closely related to the choice of boundary conditions, see
e.g.~\cite{Breitenlohner:1982jf} for an early reference.
This issue is conveniently analyzed in terms of the hypersurface inner product on the space of solutions of
the noncommutative Dirac equation, which we compute in the following.
The resulting inner product space can then be quantized by following the CAR-construction outlined in
Section~\ref{sec:quantization}.
Following the strategy developed in~\cite{Zuckerman} we consider variations of the action
functional~\eqref{eqn:actionads} and derive a~conserved current.
Explicitly, we obtain for the current density of two solutions $\psi_1$ and $\psi_2$
\begin{gather*}
iJ^\mu=i\overline{\psi_1}\hat\gamma^\mu\psi_2\sqrt{\vert g\vert}+\delta^\mu_i\mathcal{T}^{ij}
\frac{\lambda}{2}z\vartheta^\prime(z)\sqrt{\vert g\vert}
\big(\overline{\psi_1}\gamma_3\partial_j\psi_2-\overline{\partial_j\psi_1}\gamma_3\psi_2\big),
\end{gather*}
where $\sqrt{\vert g\vert} = z^{-4}$ is the square root of the metric determinant.
Since $\nabla_\mu J^\mu =0$ whenever \mbox{$\slashed{\mathrm{D}}\psi_1 = 0$} and $\slashed{\mathrm{D}}\psi_2=0$,
$J^\mu$ is a~conserved density.
We integrate it over a~f\/ixed-time hypersurface~$\Sigma$ with normal vector f\/ield $n^\mu =(1,0,0,0)^\mu$
to obtain the hypersurface inner product
\begin{gather}
\label{eqn:adsinnerprod}
(\psi_1,\psi_2)=\int_\Sigma\left(\psi_1^\dagger\psi_2+i\mathcal{T}^{0j}\frac{\lambda}{2}
z\vartheta^\prime(z)\big(\overline{\partial_j\psi_1}\gamma_3\psi_2-\overline{\psi_1}
\gamma_3\partial_j\psi_2\big)\right)\mathrm{vol}_\Sigma.
\end{gather}
That inner product is conserved only up to boundary terms, which can not be assumed to vanish on AdS.
Demanding actual conservation then yields the admissible boundary conditions.
A well-motivated restriction on the deformation is to demand $\mathcal{T}^{0i}=0$, in which case the
deformation is purely in the spatial part and no higher-order time derivatives are introduced.
In that case the hypersurface inner product~\eqref{eqn:adsinnerprod} coincides with the undeformed one,
i.e.\ $( \psi_1,\psi_2)=\int_\Sigma\psi_1^\dagger \psi_2\mathrm{vol}_\Sigma$.
We would like to stress that the solutions of the noncommutative Dirac equation are still af\/fected by the
deformation, and it would hence be of interest to study the ef\/fect of dif\/ferent choices of $T_\alpha^i$
and $\vartheta(z)$.
A natural choice would for example be such that $\mathcal{T} = \bigtriangleup$ is the spatial Laplacian on
the hypersurfaces of constant $z$, as discussed in~\cite{Schenkel:2010sc} for the Klein--Gordon f\/ield.
This choice can be implemented by taking the vector f\/ields~\eqref{eqn:rsvectors} to be parallel,
$T_{2n}^i = T_{2n-1}^i$ for all $n$.
The resulting noncommutative Dirac operator~\eqref{eqn:diracads} is still deformed, despite the vanishing
$\star$-commutation relations of the coordinates~\eqref{eqn:comads}.
This shows that our noncommutative Dirac operator does not only depend on the deformed algebra of
functions, but also on the deformed dif\/ferential calculus, which in this case is not equal to the
classical de Rham calculus (see also~\cite{Schenkel:2010zi}).

\appendix \section{Clif\/ford algebra conventions}
\label{app:gamma-conv}

For the $4$-dimensional Clif\/ford algebra we use the gamma-matrix conventions of~\cite{Aschieri:2009ky},
which are
\begin{gather*}
\eta_{ab}=\mathrm{diag}(1,-1,-1,-1)_{ab},\qquad
\{\gamma_a,\gamma_b\}\phantom{}=2\eta_{ab},\qquad
\gamma_{ab}:=\frac{1}{2}[\gamma_a,\gamma_b],
\\
\gamma_5:=i\gamma_0\gamma_1\gamma_2\gamma_3,\qquad
\gamma_5^2\phantom{}=1,\qquad
\epsilon_{0123}\phantom{}=-\epsilon^{0123}=1,
\qquad
\gamma_a^\dagger\phantom{}=\gamma_0\gamma_a\gamma_0,\qquad
\gamma_5^\dagger\phantom{}=\gamma_5.
\end{gather*}
For the $D=2$ Clif\/ford algebra we use $\eta_{ab} = \mathrm{diag}(1,-1)_{ab}$ and the Clif\/ford relation
$\{\gamma_a,\gamma_b\} = 2\eta_{ab}$ is satisf\/ied by the $2\times 2$-matrices
\begin{gather*}
\gamma_0=
\begin{pmatrix}
0&1
\\
1&0
\end{pmatrix},
\qquad
\gamma_1=
\begin{pmatrix}
0&-1
\\
1&0
\end{pmatrix},
\qquad
\gamma_3:=\gamma_0\gamma_1.
\end{gather*}
We note that $\gamma_{ab}:=\frac{1}{2}[\gamma_a,\gamma_b] = \epsilon_{ab}\gamma_3 $, where
$\epsilon_{ab}$ is the 2-dimensional $\epsilon$-tensor with $\epsilon_{01}=1$.
We further have $\gamma_a^\dagger = \gamma_0\gamma_a\gamma_0$, $\gamma_3^\dagger = \gamma_3$,
$\gamma_3^2=1$ and $\gamma_3 \gamma_a = -\gamma_a\gamma_3$.

\subsection*{Acknowledgements} We would like to thank the referees for their constructive and useful
comments.
CFU is supported by Deutsche Forschungsgemeinschaft through the Research Training Group GRK\,1147
\textit{Theoretical Astrophysics and Particle Physics}.

\pdfbookmark[1]{References}{ref}
\LastPageEnding

\end{document}